\pdfoutput=1
\documentclass{aa} 
\usepackage[varg]{txfonts}
\bibpunct{(}{)}{;}{a}{}{,} 

\usepackage[colorlinks,linkcolor=blue,
           anchorcolor=blue,citecolor=blue
           ]{hyperref}
\usepackage{natbib}
\usepackage{amsmath}
\usepackage{amssymb}
\usepackage{url}
\usepackage{longtable}
\usepackage{multirow}

\usepackage{graphicx}   
\usepackage{epsfig}
\usepackage{bm}

\renewcommand{\vec}[1]{ {\mathbf #1} }

\newcommand{\Fig}{{Figure}}

\newcommand{\divB}{\nabla\cdot\mathbf{B}}
\newcommand{\crlB}{\nabla\times\mathbf{B}}

\newcommand{\RNum}[1]{\uppercase\expandafter{\romannumeral #1\relax}}

\begin{document}

   \title{The Initiation Mechanism of the First On-disk X-Class Flare of Solar Cycle 25}

   \author{Aiying Duan\inst{\ref{inst1}, \ref{inst2}}
          \and Chaowei Jiang\inst{\ref{inst3}}
          \and ZhenJun Zhou\inst{\ref{inst1}}
          \and Xueshang Feng\inst{\ref{inst3}}}

   \institute{Planetary Environmental and Astrobiological Research Laboratory (PEARL),
              School of Atmospheric Sciences, Sun Yat-sen University, Zhuhai
              519000, China\\
              \email{duanaiy@mail.sysu.edu.cn}\label{inst1}
      \and Yunnan Key Laboratory of the Solar physics and Space Science, Kunming 650216, China\label{inst2}
      \and Institute of Space Science and Applied Technology, Harbin Institute of Technology, Shenzhen 518055, China\label{inst3}
              }

 \abstract{ In this paper we study the initiation mechanism of the first on-disk X-class eruptive flare in solar cycle 25. Coronal magnetic field reconstructions reveal a magnetic flux rope (MFR) with configuration highly consistent with a filament existing for a long period before the flare, and the eruption of the whole filament indicates that the MFR erupted during the flare. However, quantitative analysis
 shows that the pre-flare MFR resides in a height too low to trigger a torus instability (TI). The filament experienced a slow rise before the flare onset, for which we estimate evolution of the filament height using a triangulation method by combining the SDO and STEREO observations, and find it is also much lower than the critical height for triggering TI.
  On the other hand, the pre-flare evolution of the current density shows progressive thinning of a vertical current layer on top of the flare PIL, which suggests that a vertical current sheet forms before the eruption. Meanwhile, there is continuously shearing motion along the PIL under the main branch of the filament, which can drive the coronal field to form such a current sheet. As such, we suggest that the event follows a reconnection-based initiation mechanism as recently established using a high-accuracy MHD simulation, in which an eruption is initiated by reconnection in a current sheet that forms gradually within continuously-sheared magnetic arcade. The eruption should be further driven by TI as the filament quickly rises into the TI domain during the eruption.}

   \maketitle

\section{Introduction}
\label{sec:intro}
Solar eruptions, such as solar flares and coronal mass ejections (CMEs), are catastrophic energy-conversion phenomena which can cause severe space weather that impacts heavily on advanced technological systems in modern society. It is well established that the solar magnetic field plays a fundamental role in such eruptions~\citep{Forbes2006, ChenP2011, Shibata2011, ChengX2017, GuoY2017}, thus it is important to understand the pre-eruptive structure of the magnetic field for a reliable prediction of solar eruptions. In addition, solar eruption is believed to be a drastic release of the free energy stored in the complex, unstable magnetic field in solar corona, thus understanding the initiation mechanism of eruption is also crucial for a physics-based forecast of eruptions~\citep{Kusano2020}.

A particular magnetic structure that holds the central position in study of solar eruptions is the magnetic flux rope~\citep[MFR,][]{Kuperus1974, ChenJ1989, Titov1999, Amari2014nat}. An MFR refers to a coherent group of twisted magnetic flux winding around a central axis, and is believed to exist commonly in the corona as well as in the whole heliosphere. It is of great interest to the study of eruption initiation owing to the two well-known ideal magnetohydrodynamics (MHD) instabilities of the MFR, namely the kink instability~\citep[KI,][]{Hood1981, Torok2004, Torok2010, Torok2005} and the torus instability~\citep[TI,][]{Kliem2006}, which can result in a sudden loss of equilibrium of a pre-eruption MFR and lead to eruptions. The twist degree of the MFR (denoted by $\mathcal{T}$, i.e., the winding number of magnetic field lines around the axis) and the decay index of the strapping field ($n$, quantifying the spatial decreasing speed of the overlying magnetic field) are controlling parameters of the KI and TI of MFR, respectively. Evolution of the photosphere slowly drives an MFR to the unstable regime when these control parameters exceed their critical values, and an eruption will be triggered through these instabilities.

Since these ideal MHD models of solar eruptions are based on MFR, once there is evidence indicating the existence of MFR prior to an eruption, it is often considered that the eruption could be triggered by the MFR instabilities. Indeed, our recent survey of the major flares (above GOES M5.0) from 2011 to 2017 with coronal magnetic field extrapolations from vector magnetograms taken by the \emph{Helioseismic and Magnetic Imager} (HMI) onboard \emph{Solar Dynamics Observatory}~\citep[SDO][]{Pesnell2012} shows nearly $90\%$ of the events possess pre-flare MFRs~\citep{DuanA2019, DuanA2021apjl}, which indicates that the pre-existence of MFR is rather common for major flares. Furthermore, by calculating the controlling parameters $T_w$ and $n$ for each of these events, it is found that many events have the parameters exceeding their thresholds for triggering the MFR instabilities. However, there are over half of MFR-possessing events cannot be explained fully by the MFR instabilities, since for these events both the two parameters before flare are apparently below their thresholds for triggering the MFR instabilities.

In this paper, we present a study of the first on-disk X-class flare of solar cycle 25, for which there are clear evidences for a pre-existing MFR but it is not likely the MFR instabilities that triggered the eruption, as supported by both extrapolations of coronal magnetic field and analysis of time profile of the rising filament height. We further found that the pre-flare evolution of the current density shows progressive thinning of a vertical current layer on top of the flare PIL, which suggests that a vertical current sheet forms before the eruption as driven by the photospheric motions. This hints that the reconnection may have triggered the eruption.
Study of the filament height evolution suggests that once being triggered, the MFR immediately runs into the domain of TI and should be also driven by TI in addition to the flare reconnection.

\section{Data and Method}
\label{sec:data}
Data from the SDO are mainly used in this study. The \emph{Atmospheric Imaging Assembly}~\citep[AIA,][]{Lemen2012} on board SDO provides seven EUV filtergrams that cover the temperature from $10^5$~K to $10^7$~K, with a spatial resolution of $1.2$~arcsec (i.e., two pixels) and a cadence of 12~s.
The H$\alpha$ data from the Global Oscillation Network Group (GONG) are also used to show the filament with a spatial resolution of 1~arcsec. In addition, the EUVI images from STEREO-A is used to determine the height of the filament around the eruption onset time by triangulation with the AIA images.

The HMI~\citep{Schou2012} instrument on SDO provides vector magnetograms with different products, and one of them is the Space-weather HMI Active Region Patch~\citep[SHARP,][]{Bobra2014} data which is suitable for extrapolation in Cartesian coordinates. The SHARP data has effective spatial and temporal resolutions of $\sim 1$~{arcsec} and 12~min, respectively. It has been resolved with 180$^\circ$ ambiguity using the minimum energy method and transformed into a heliographic Cylindrical Equal-Area (CEA) projection centered on the patch via the Lambert method with the projection effect corrected.

We use the CESE--MHD--NLFFF extrapolation code developed by~\citet{Jiang2013NLFFF} to reconstruct the coronal magnetic field from the photospheric field as provided by the SHARP data. The raw data are preprocessed~\citep{Jiang2014Prep} to reduce the noise and the Lorentz force before being put into the code. The CESE--MHD--NLFFF code uses an advanced space-time conservation-element and solution-element (CESE) scheme~\citep{Jiang2010} to solve a set of modified MHD equations with a friction force and a zero$-\beta$ approximation, seeking approximately force-free equilibrium for a given vector magnetogram based on an MHD-relaxation approach. It is widely used in reconstructing different magnetic structures corresponding to the SDO observations~\citep{Jiang2013NLFFF, Jiang2014Prep, DuanA2017, ZouP2019, ZouP2020, DuanA2021apj, DuanA2021apjl, DuanA2022AA} and shows a good quality of extrapolations with reasonably small metrics that measure the force-freeness and divergence-freeness of the field.

To identify MFRs in the extrapolated field, we calculate the magnetic twist number $T_w$~\citep{Berger2006} with a fast code proposed by~\citet{LiuR2016}. For a given magnetic field line, $T_w$ is defined as
\begin{equation}\label{Tw}
  T_w=\int_L \frac{(\nabla \times \vec B)\cdot \vec B}{4\pi B^2} dl,
\end{equation}
where the integral is taken along the field line between its two footpoints, both of which are anchored at the photosphere. We noted that $T_w$ is not the exact number of turns that field lines enwind a common axis, but it can provide a good approximation of two closely neighboring field lines winding about each other, and the particular field line with the maximum local twist number $|T_w|_{max}$ can be taken as a reliable proxy of the MFR axis~\citep{LiuR2016, DuanA2019}.

Since $|T_w|_{\max}$ merely measures the twist at an infinitesimal radius around the MFR axis, a measurement of the large-scale twist rather than only a local twist should be used to a more comprehensive judgement of KI. Thus, we use two methods to quantify the large-scale twist in the paper as we did in~\citet{DuanA2022AA}. In the first way, a flux-weighted average of the twist number was defined as
\begin{equation}
(T_w)_{\rm mean}= \frac{\int_{S, T_w \ge 1} B_n T_w ds}{\int_{S, T_w \ge 1} B_n ds},
\end{equation}
where $S$ and $B_n$ denote a cross section of the MFR and the magnetic field component normal to the cross-section slice, respectively. We restrict the integration with the area of $T_w \ge 1$ since the MFR is defined by this threshold and the studied MFR here has positive twist numbers. In the second method, we calculate the winding number directly for each field line (denoted by $\mathcal{T}$, which is different from $T_w$) around the axis using the standard equation of~\citet{Berger2006},
\begin{equation}
\mathcal{T}= \frac{1}{2\pi }\int_L \mathbf{T}(l)\cdot \mathbf{V}(l) \times \frac{d\mathbf{V}(l)}{dl} dl,
\end{equation}
where $l$ is the arc length from a reference starting point on the axis curve, $\mathbf{T}(l)$ is the unit vector tangent to the axis and $\mathbf{V}(l)$ denotes a unit vector normal to $\mathbf{T}(l)$ and pointing from the axis curve to the target field line. To give a reasonable estimate of the average value of the winding number, i.e., the large-scale twist, we also calculated the flux-weighted average of the winding number,
\begin{equation}
\mathcal{T}_{\rm mean}= \frac{\int_{S, T_w \ge 1} B_n \mathcal{T}ds}{\int_{S, T_w \ge 1} B_n ds}.
\end{equation}

To evaluate the condition of TI, we calculate the decay index of the strapping field of the MFR. It is defined as
\begin{equation}\label{n1}
  n_z=-\frac{d \log(B)}{d \log(z)},
\end{equation}
where $B$ and $z$ denote the strapping magnetic field and the radial distance (or height) from the solar surface, respectively. Apparently, this formula is valid only for those MFRs which erupt radially (i.e., vertically) and have strapping force in the opposite radial direction. However, non-radial filament eruptions are often observed~\citep{McCauley2015}, some of them even deviate from the radial direction significantly. Thus, we recommend a more accurate way to calculate the decay index, which was first proposed in~\citet{DuanA2019}. Briefly speaking, the decay index $n$ is computed along an oblique line matching the eruption direction as
\begin{equation}\label{n2}
  n=-\frac{d\log(B_{p})}{d\log(r)}.
\end{equation}
Here, $r$ is the distance between the surface and the apex of the MFR's axis along the eruptive direction. $B_p$ refers to the poloidal flux, the product between the current of the rope and $B_p$ can produce an effective strapping force in the opposite direction of the eruption. It is decomposed from the magnetic strapping field which is extrapolated with the potential field model. One can find detailed description of the method in~\citet{DuanA2019}. In this paper, we use both of the methods (i.e., Equations \ref{n1} and \ref{n2}) to calculate the decay index.

\begin{figure*}
  \centering
  \includegraphics[width=0.9\textwidth]{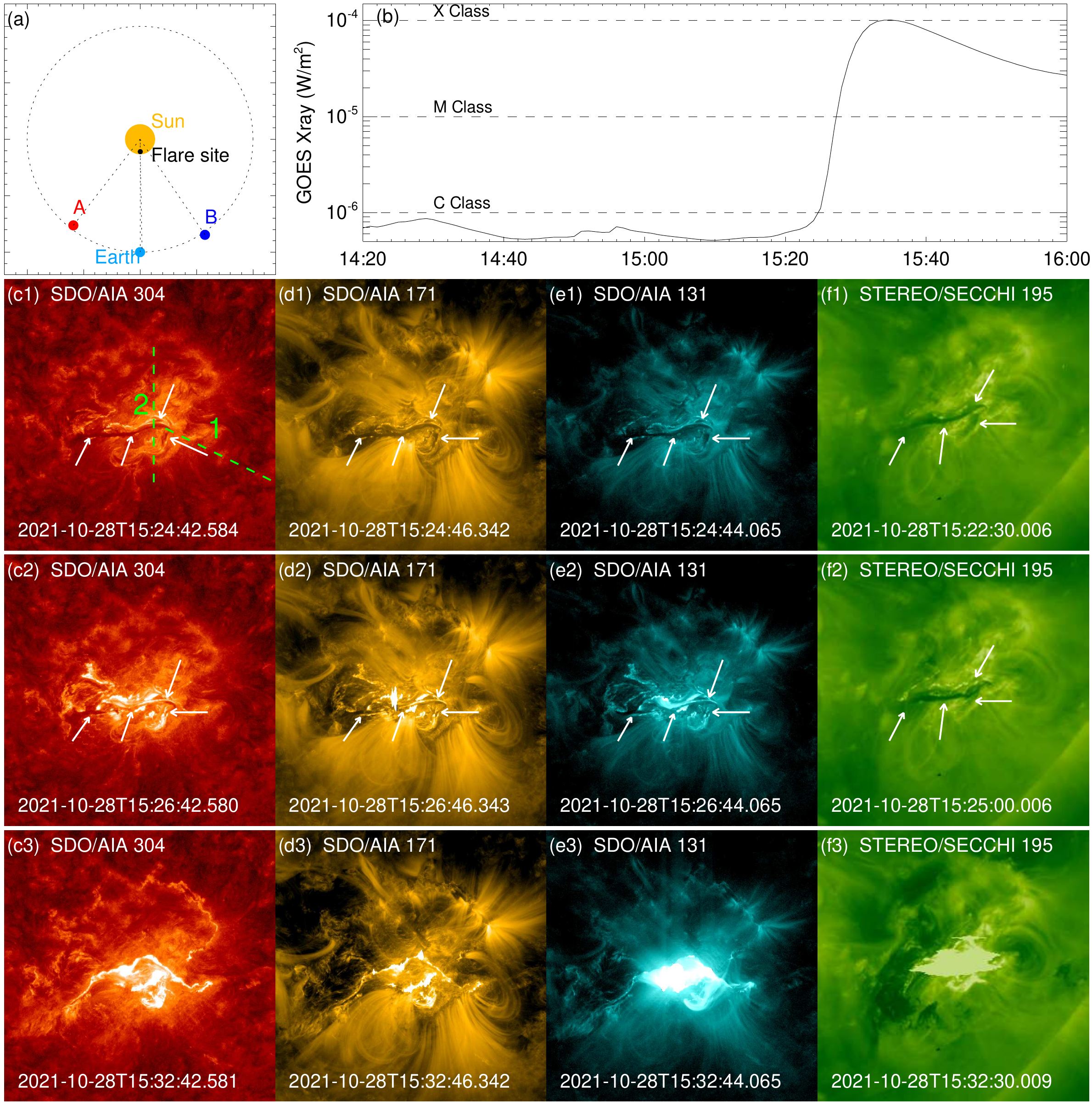}
  \caption{(a) Angular locations of the Sun, the flare site, and STEREO, with respect to the Earth. (b) GOES soft X-ray flux. (c)-(f) Observations of the filament motion during the eruption from SDO/AIA and STEREO-A/EUVI with wavelengths of 304~{\AA}, 171~{\AA}, 131~{\AA}, and 195~{\AA}, repectively. The first row display a clear J-shaped filament which remains stable before the flare onset. The second row show that the filament starts to rise up quickly. The third row present the eruption of the filament and the flare ribbons. An animation of SDO/AIA observation with cadence of 1~minute is attached.}
  \label{f1}
\end{figure*}

\begin{figure*}
  \centering
  \includegraphics[width=0.8\textwidth]{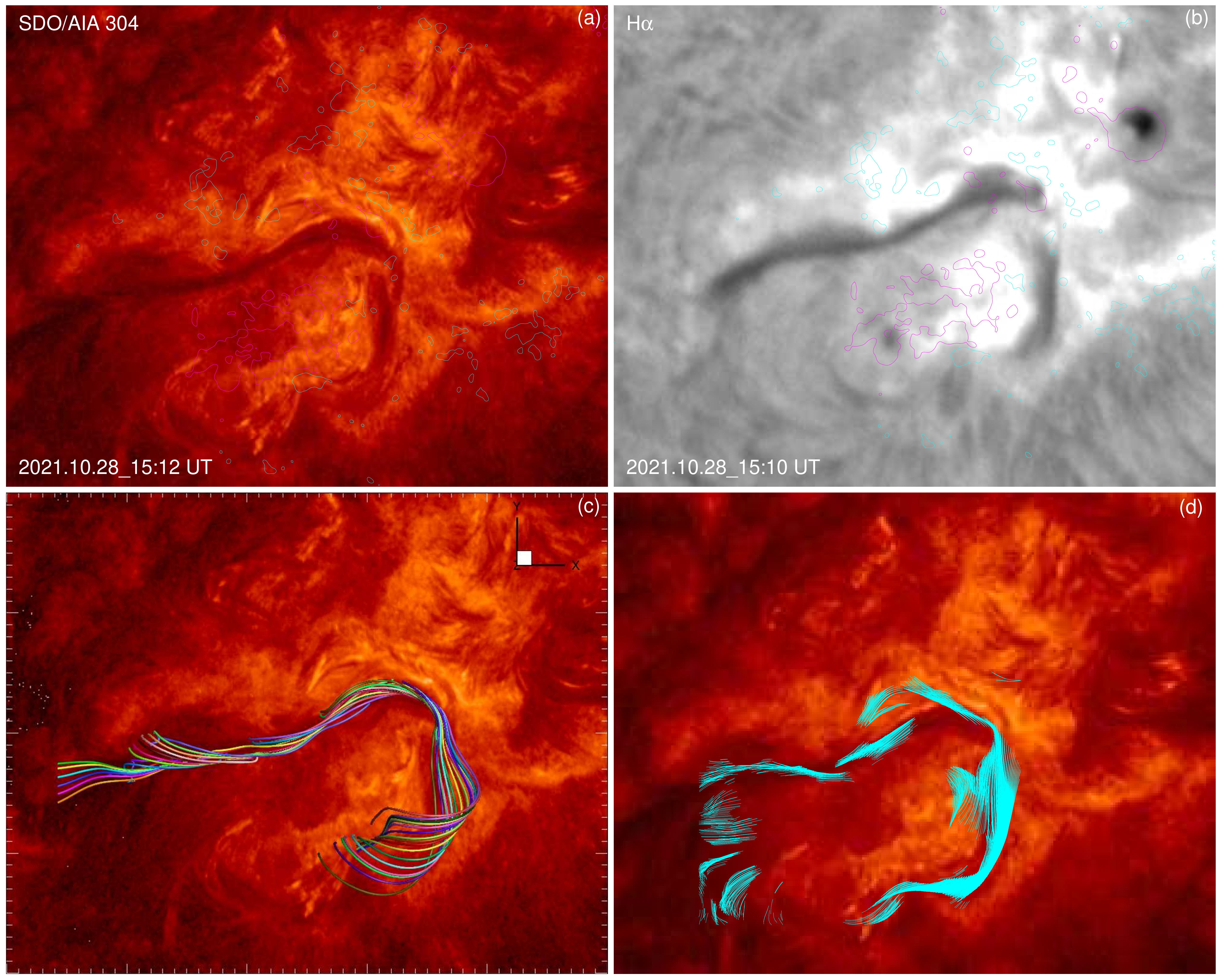}
  \caption{Observations of the filament and the NLFFF reconstruction of the MFR. (a) SDO/AIA 304~{\AA} image at 15:12 UT. (b) H$\alpha$ image from Cerro Tololo Inter-American Observatory at 15:10 UT. (c) Sampled field lines (with different colors) from the reconstructed magnetic  field at 13:24 UT. (d) Segments of magnetic field lines that extend from magnetic dips to a scale height of 300~km (the cyan lines) at 13:24 UT.}
  \label{f2v1}
\end{figure*}

\section{Results}
\label{sec:res}

\subsection{Overview of the event}
The active region NOAA 12887 (i.e., AR 12887) rotated from the back side of the Sun to the disk on 2021 October 22, and produced many flares during its passage on the disk in the following days. In particular, when it was close to the central meridian on 2021 October 28 (located at S28W01 on the solar disk), the AR had developed to a complex $\beta\gamma$-type sunspots system, and generated multiple flares, including five C-class ones, two M-class ones (an M1.4 and M2.2 peaking at 07:40~UT and 10:28~UT, respectively), and the largest one of X1.0 (peaking at 15:35~UT), which is of our interest. This X1.0 flare is also the first X-class flare observed on solar disk in solar cycle 25 as an eruptive flare accompanied with a CME.

\Fig~\ref{f1} (and its supplementary animation) presents the SDO and STEREO-A observations of the flare eruption in different EUV channels. As seen in the AIA 304~{\AA} (denoted by the arrows in \Fig~\ref{f1}c) and also the H$\alpha$ (\Fig~\ref{f2v1}b) images, there was clearly a J-shaped filament existing well before the flare. The filament existed without significant variation at least from the beginning of the day and suggests the pre-existence of a stable MFR. This filament run roughly along the main polarity inversion line (PIL) (i.e., the core site where the flare took place), and the X1.0 flare was associated with the eruption of this J-shaped filament system. The flare initiated at 15:22~UT, reaching its peak impulsively within about 10 minutes, and the flare ribbons expanded along the filament towards both the east and south directions. Immediately after onset of the flare, the middle of the straight part of the J-shaped filament first rose rapidly, which was followed by eruption of the whole filament (as denoted by arrows in middle panels of \Fig~\ref{f1}) with dark materials ejected fast from the flare site (see the animation for whole eruption). This eruption resulted in a CME with a wide angular width of around $270^{\circ}$ and an average speed of 502 km s$^{-1}$, which was detected by SOHO/LASCO C2 coronagraph at 15:48 UT~\footnote{http://helio.gmu.edu/seeds/lasco.php}. We note that during the eruption, there was no obvious kink motion (i.e., with a helical shape) of the erupting filament, suggesting that the eruption is not likely triggered by the KI.

\subsection{Coronal magnetic flux rope}
\begin{table*}[htbp]
  \centering
  \caption{Metrics of force-freeness and divergence-freeness for all the extrapolated coronal fields.}
    \begin{tabular}{ccccc}
    \hline
    \hline
    Area & CWsin & $\langle |f_{i}|\rangle$  & $E_{\crlB}$ & $E_{\divB}$ \\
    \hline
    Whole Region & $0.30\pm0.007$ & $(4.2\pm0.2)\times 10^{-4}$ & $0.17\pm0.01$ & $(3.3\pm0.02)\times 10^{-2}$ \\
    MFR Region &  $0.27\pm0.006$ & $(1.4\pm0.09)\times 10^{-3}$ & $0.17\pm0.004$ & $(8.0\pm0.6)\times 10^{-2}$ \\
    \hline
  \end{tabular}
  \label{tab:1}
\end{table*}

To confirm the pre-existence of the MFR, we extrapolate the pre-flare coronal magnetic fields of a time series from 13:00~UT to 15:12~UT with time cadence of 12 minutes. The quality of force-freeness and divergence-freeness of the extrapolated field has been checked with two groups of metrics. One group is named as CWsin and $\langle |f_{i}|\rangle$, which are commonly used in the literature~\citep[e.g.,][]{Schrijver2006, DeRosa2009, Jiang2013NLFFF}. They are, respectively, mean sine of the angle between current density $J$ and magnetic field $B$ as weighted by $J$, and the normalized divergence error. The other is $E_{\crlB}$ and $E_{\divB}$, which are, respectively, the average ratios of the residual Lorentz force and the nonphysical force ($F = B\divB$, as introduced by the non-zero $\divB$) to the sum of the magnitudes of the two component forces (i.e., the magnetic tension force and magnetic pressure-gradient force)~\citep{Jiang2012apj, Malanushenko2014}. Although less used than CWsin and $\langle |f_{i}|\rangle$, these two metrics are introduced for a more physically meaning by directly quantifying the residual force in the extrapolated field (one can refer to the Appendix of~\citet{DuanA2022AA} for detailed descriptions of all these metrics). Among them, the CWsin and $E_{\crlB}$ measure the force-freeness of the field, while the $\langle |f_{i}|\rangle$ and $E_{\divB}$ estimate the divergence-freeness of the field. Here for all the extrapolated fields, we calculated the average and rms values of these metrics for the whole volume and the subdomain containing only the flux rope, respectively, and the results are shown in Table~\ref{tab:1}. We can see that all of the metrics are highly consistent with our previous works~\citep{Jiang2013NLFFF, DuanA2017, DuanA2021apj, DuanA2022AA}. It should be noted that the extrapolated field is not accurately force-free, mainly because the photospheric field is generally not force-free, which conflicts with the basic assmuption of force-free field modeling. For example, the CWsin$=0.27$ implies an average misalignment angle of $16^{\circ}$ between the current density and magnetic field. This value is on the same level with that from other currently avaiable NLFFF codes for ARs, which give CWsin in a typical range of $0.2\sim0.4$~\citep[e.g.,][]{DeRosa2009, DeRosa2015, Schrijver2006, Schrijver2008a}. For the residual force, as quantified by $E_{\crlB}$ and $E_{\divB}$, it is smaller than the magnetic pressure and tension force by an order of magnitude.

\begin{figure}
  \centering
  \includegraphics[width=0.45\textwidth]{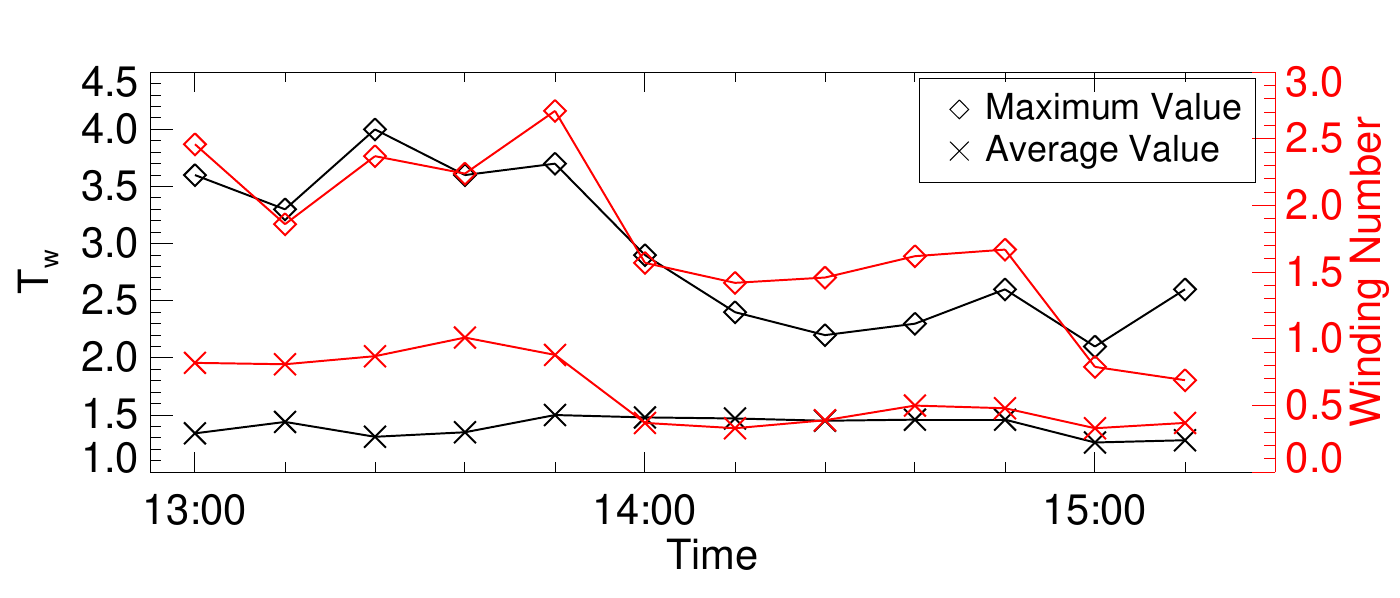}
  \caption{Evolution of the twist number $T_w$ and the winding number $\mathcal{T}$, including both their largest value and mean value.}
  \label{tww}
\end{figure}

From the reconstructed field, we found a large-scale MFR with the $T_w$ of its axis kept larger than $2.0$. To compare the reconstructed MFR with the observational filament, we show observations by SDO/AIA 304~{\AA} and H$\alpha$ from Cerro Tololo Inter-American Observatory in \Fig~\ref{f2v1}(a) and (b), respectively. Since the filament can be observed all the time before its eruption without much variation from the start of October 28, here we only display the time (15:12~UT for AIA 304~{\AA} and 15:10~UT for H$\alpha$) when the filament can be seen most clearly. \Fig~\ref{f2v1}(c) presents the sampled reconstructed magnetic field lines (the colored thick curves) at 13:24~UT. Note that at 13:24~UT, the $T_w$ of the MFR axis reaches its maximum value of $\sim 4.0$. In \Fig~\ref{f2v1}(c) the field lines are overlaid on the AIA image of panel (a), and such comparison clearly demonstrates that the MFR almost matches perfectly with the filament system. \Fig~\ref{f2v1}(d) shows the magnetic dips at time of 13:24~UT. We compute the location of dips from a 3D magnetic field according to their definition (i.e., the location with $B_z = 0$ and $\vec B\cdot \nabla B_z \ge 0$) and trace the parts of the field lines that extend from the dips up to a height of 300~km ~\citep[which is a typical scale height in prominence,][]{Guo2010, Zuccarello2016} to simulate the location of filament, and it shows that the magnetic dips highly matches with the filaments system, as the MFR does.

\Fig~\ref{tww} shows both the largest ($(T_w)_{max}$ and $\mathcal{T}_{max}$) and flux-weighted average values ($(T_w)_{mean}$ and $\mathcal{T}_{mean}$) of the twist number and winding number of the flux rope. As expected, there are systematic differences between  $T_w$ and $\mathcal{T}$. In addition, the averaged winding number that quantify the large-scale twist of the MFR are below the smallest threshold for KI, namely 1.25~\citep{Hood1981}, suggesting that KI cannot be triggered.

\begin{figure*}
  \centering
  \includegraphics[width=0.8\textwidth]{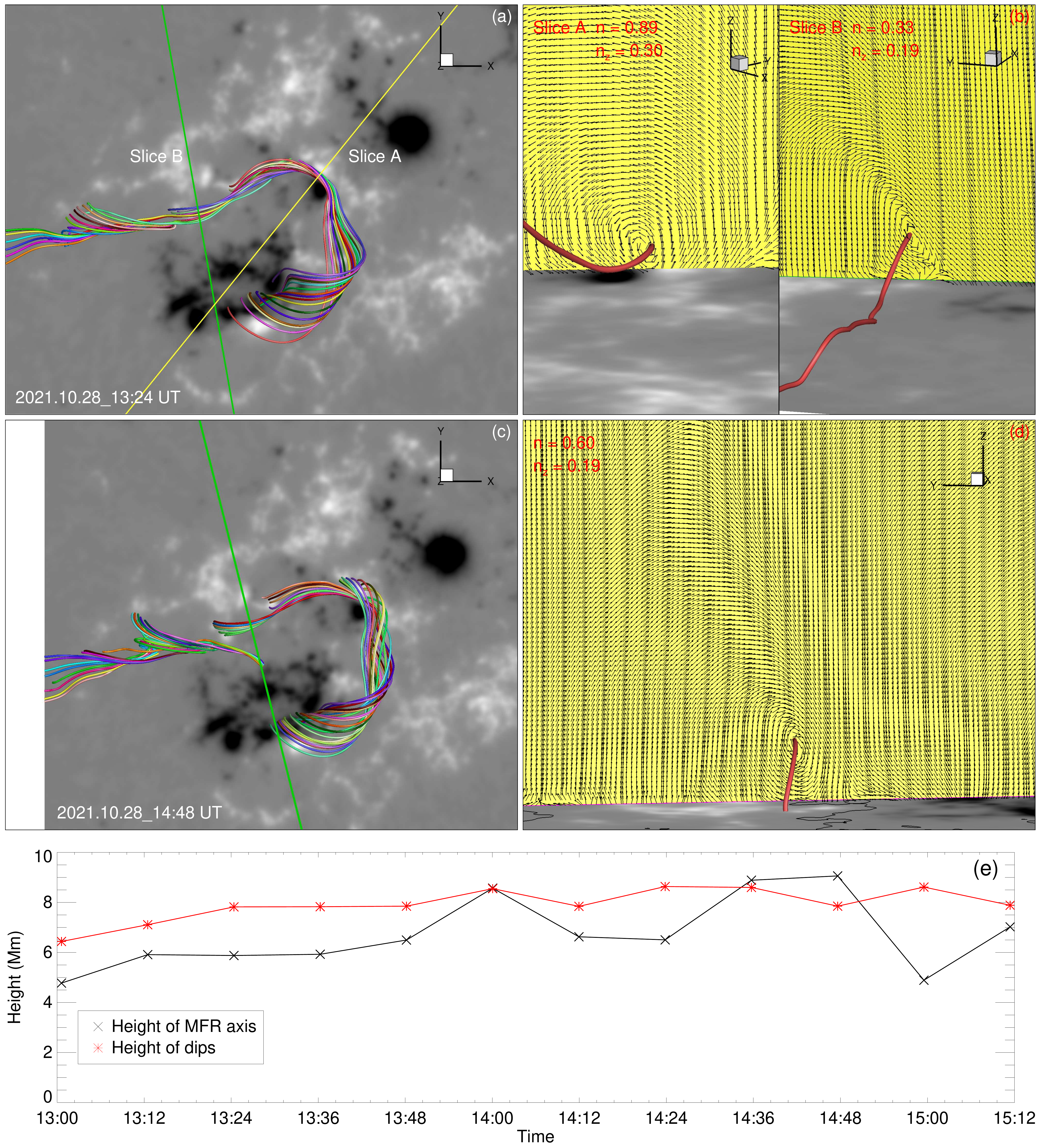}
  \caption{Observations of the filament and the NLFFF reconstruction of the MFR. (a) Sampled field lines (with different colors) from the reconstructed magnetic  field at 13:24 UT. (b) two vertical cross sections of the MFR whose locations are denoted by the yellow and green lines in panel (a). (c) Same as (a) but for 14:48 UT. (d) A vertical cross section of the MFR located at the position marked by the green line in panel (c). The arrows in both (b) and (d) show the directions of the transverse magnetic field on the slices, and they form rings centered at the red thick line which is expected as the axis of the MFR. The decay index calculated at the axis of the MFR are denoted in each slice. (e) Evolution of the heights of the MFR axis and the magnetic dips at their highest point in the section corresponding to the earliest rising part of the filament during its eruption.}
  \label{f2v2}
\end{figure*}

\Fig~\ref{f2v2}(a) and (c) present the sampled reconstructed magnetic field lines (the colored thick curves) at two representative times, 13:24~UT and 14:48~UT,  respectively, when the reconstructed MFR can be most clearly shown. \Fig~\ref{f2v2}(b) and (d) show vertical cross sections of the MFR whose locations are marked by the yellow and green lines in the panels (a) and (c), respectively. The arrows on the slice show the directions of the transverse field components, from which we can see that in both panels the arrows show the poloidal flux of the rope forms spirals centered at the axis of MFR (denoted by the thick red lines, which are also the field line in the MFR with the largest $T_w$). Values of the decay index calculated from different positions with the two methods (the radial definition $n_z$ and the oblique definition $n$) are labeled on (b) and (d). In panel (b), slice A and B correspond to the yellow and green lines on panel (a), and the decay indices are computed at the points where the yellow and green lines intersect with the MFR, respectively. Similarly, decay index labeled in panel (d) is calculated at the point of the intersection between the green line and the MFR on panel (c). Here we have $n$ ($n_z$) with $0.89$ $(0.30)$, $0.33$ $(0.19)$ and $0.60$ $(0.19)$ for the three points, while the corresponding heights of the MFR's axis are $3.33$, $8.00$ and $12.43$ with unit of arcsecond (or 720 km), separately. Such low values of decay index at the MFR's axis are far below the canonical threshold ($\sim 1.5$) of TI, and also below the most typical range of $1.5 \pm 0.2$ as derived in many theoretical, numerical, and laboratorial investigations ~\citep{Kliem2006, Fan2007, Torok2007, Aulanier2010, Demoulin2010, Fan2010, Myers2015, Inoue2014, Inoue2018, McCauley2015, Zuccarello2015, Alt2021} as well as some statistical studies of flare events~\citep{DuanA2019, DuanA2021apjl}.

\Fig~\ref{f2v2}(e) further shows the evolution of the heights of MFR axis and magnetic dips from 13:00~UT to 15:12~UT. The MFR axis is identified using exactly the same method as done in~\citet{DuanA2019}, i.e., it is the field line in the MFR with the largest twist number $T_w$. Then we tracked the highest point of the axis in the section corresponding to the earliest rising part of the filament during its eruption. As can be seen, the height increases with time overall but from 13:48~UT to 15:12~UT it shows oscillation from around 5 to 9~Mm, which is likely owing to the uncertainty of the NLFFF extrapolations (of $\sim \pm 2$~Mm). The magnetic dips, which are calculated from the NLFFF data, have the height of around 8~Mm on average. As is often assumed, the filament material is supported by magnetic dips, and thus the height of the dips can be taken as an approximation of the upper limit of the filament height. All these heights at all time are far below the TI critical height of $\sim 25$~Mm (see in \Fig~\ref{f5} the profile of the decay index of the strapping field), suggesting that TI is not likely the trigger mechanism of eruption in this event. It is worth noting that at some times, the dips are slightly higher than the MFR axis by around 2~Mm.  This is owing to the complexity of the MFR axis, which is not simply arched (as shown in the idealized model of MFR, for example,~\citet{Zuccarello2016}) but undulated and thus is itself dipped at some locations (see \Fig~\ref{f2v2}(b)).

\subsection{Kinematics of the erupting features}

\begin{figure*}
  \centering
  \includegraphics[width=0.6\textwidth]{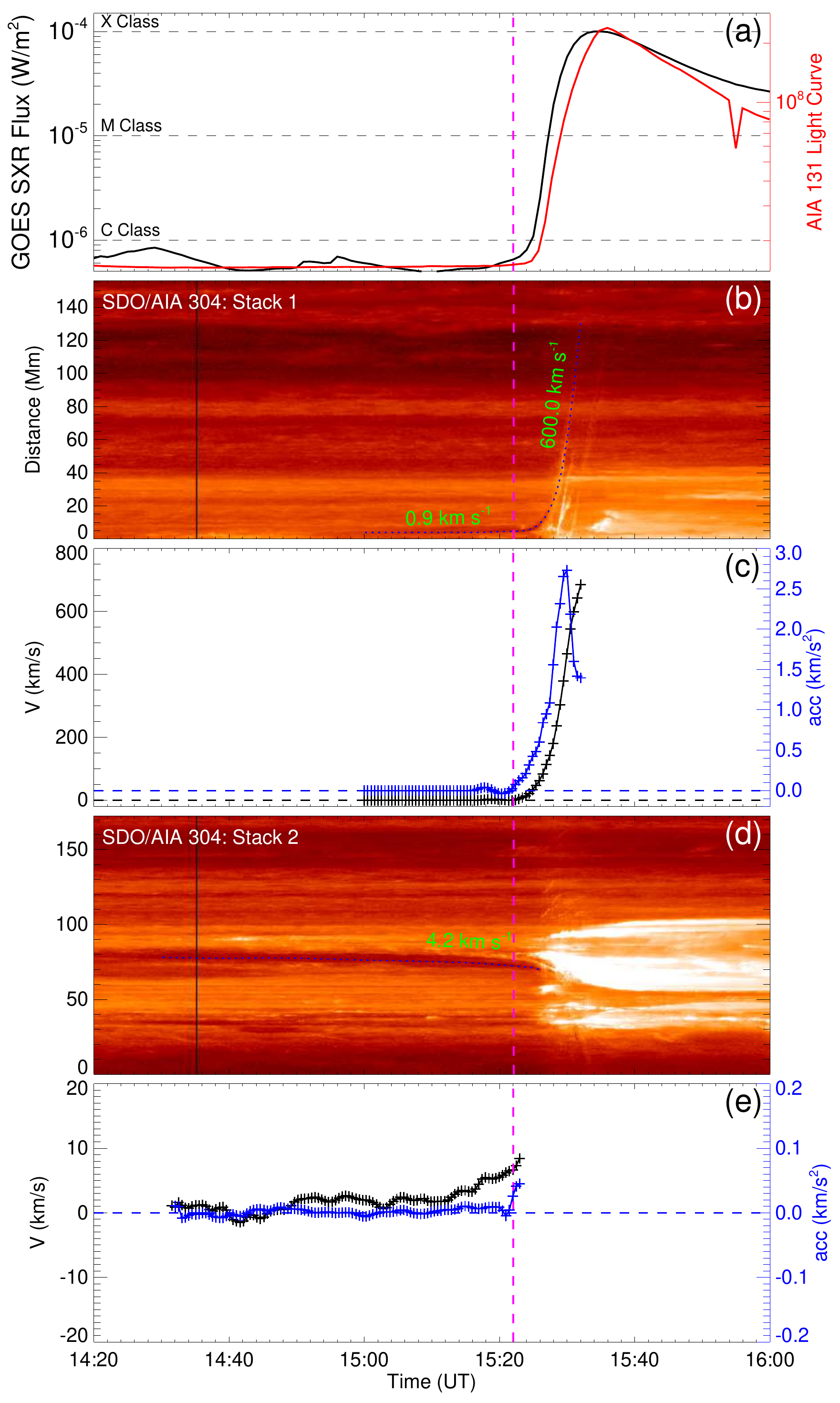}
  \caption{(a) Evolutions of the GOES soft X-ray flux (dark solid line) and the source-integrated AIA 131~{\AA} flux (red solid line) from 14:20~UT to 16:00~UT. (b) Stack plot (or Jmap) of the 304~{\AA} intensities along slit 1 as denoted in \Fig~\ref{f1}(c1), showing the fast rise of the filament. The dashed blue line tracks the leading edge of the erupting filament. (c) Temporal evolutions of the velocity and acceleration of the eruptive filaments as computed from the dashed blue line in panel (b). (d) Stack plot along slit 2 as denoted in \Fig~\ref{f1}(c1), which tracks mainly the slow rise of the filament before the eruption onset. (e) The velocity and acceleration for the slow rise of the filament are estimated from the dashed blue line in panel (d). In all the panels, the vertical dashed line colored with pink denotes the time of 15:22~UT.
  An animation for the SDO/AIA 171~{\AA} and 304~{\AA} with a cadence of 12~second is attached (without the light curves).}
  \label{f3}
\end{figure*}

We further analyze the kinematic evolutions of the eruptive filament and its overlying loops, as well as their timing relations with the evolutions of the flare emission, and the results are shown in \Fig~\ref{f3} (and the supplementary animation). Specifically, \Fig~\ref{f3}(a) presents the GOES soft X-ray flux (the black line). Regarding that the GOES soft X-ray flux is taken for the full disk and may include a contribution from other regions, we also plot the source-integrated AIA 131~{\AA} flux to represent the hot flare emission as restricted within the studied AR. These light curves confirm that the onset time of the flare impulsive phase (or fast reconnection) is 15:22~UT (as denoted by the pink dashed line).

\Fig~\ref{f3}(b) and (c) show time-distance evolution and the estimated speed (and acceleration) in slit 1 from the 304~{\AA} as denoted in \Fig~\ref{f1}(c1), which is selected to track most clearly the fast ejection of the filament material. In addition, \Fig~\ref{f3}(d) and (e) show those for slit 2 as denoted in \Fig~\ref{f1}(c1) to track mainly the slow rise of the filament before the eruption onset.
The filament undergoes a slow rise from around 14:40~UT to the flare onset (15:22~UT) with speed increased mildly from almost zero to a few km~s$^{-1}$ (note that the acceleration is very small of around $0.01$~km~s$^{-2}$, see \Fig~\ref{f3}e), which is close to the quasi-static evolution speed as driven by the photosphere (e.g., smaller than the photospheric sound speed). After 15:22~UT, the filament was impulsively accelerated to a speed of $\sim 600$~km~s$^{-1}$ within around 8 minutes (\Fig~\ref{f3}c). Thus, the start time of the filament eruption is in accordance with the abrupt increasing of both the GOES soft X-ray and the AIA 131~{\AA} light curves.

\begin{figure*}
  \centering
  \includegraphics[width=0.9\textwidth]{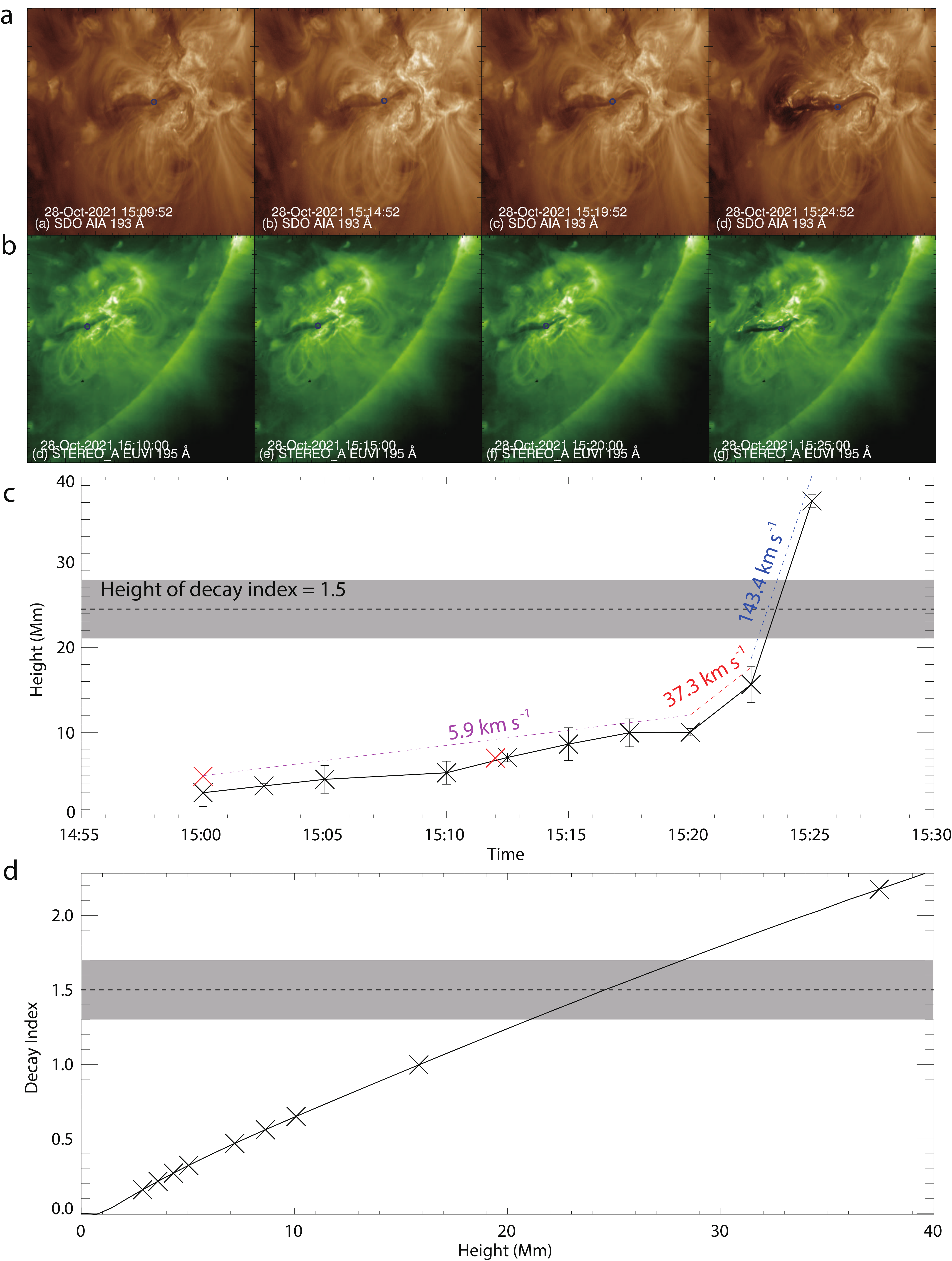}
  \caption{Height evolution of the filament as determined by triangulation method combining SDO/AIA (a) with STEREO/EUVI (b) images. The highest point of the filament is denoted by the small circles in both the AIA and EUVI images. (c) Evolution of the height and the typical velocity. The uncertainty in the estimation of the height is denoted by the error bars. The two red crosses denote the heights of the MFR derived from the NLFFF extrapolations for time of 15:00~UT and 15:12~UT, respectively. The dashed lines shows the critical height of TI using the canonical threshold ($1.5$) of decay index, and the gray region shows range of height with decay index from $1.3$ to $1.7$. (d) Profile of the decay index of the strapping field. The heights of the filament as shown in (c) are also denoted on the profile by the X symbols.}
  \label{f5}
\end{figure*}

\begin{figure*}
  \centering
  \includegraphics[width=\textwidth]{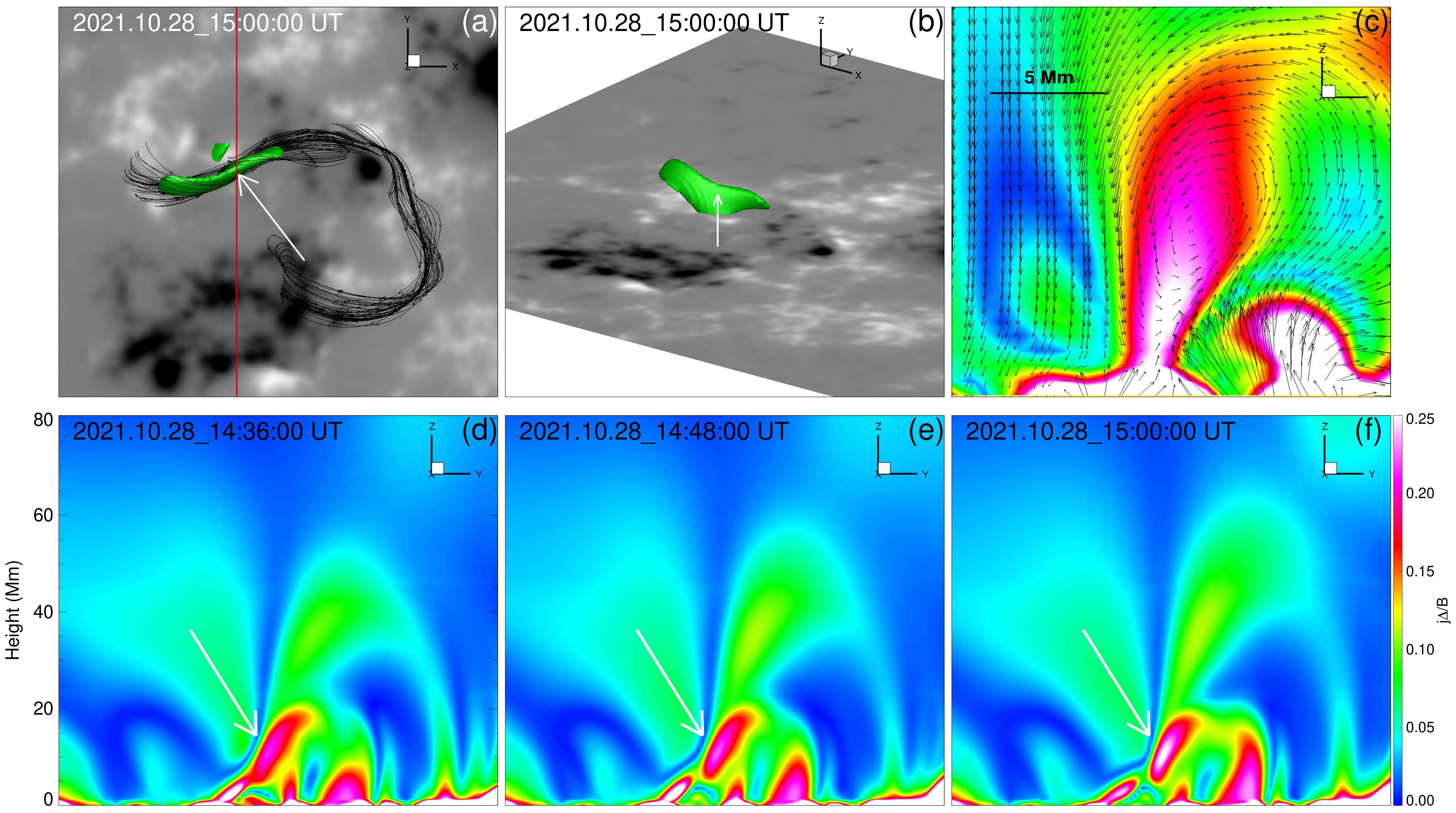}
  \caption{Thinning of a vertical current layer as derived from the NLFFF extrapolations. (a) The green object shows the 3D iso-surface of $j\Delta/B=0.2$, i.e., a thin current layer on top of the flare PIL. The black lines show sampled magnetic field lines of the MFR, and note that most of these field lines cross through the current layer. (b) A 3D view of the current layer. (c) A vertical slice cutting perpendicularly through the current layer at the largest value of $j\Delta/B$. The arrows show magnetic field vectors on the slice. (d)-(f) Time evolution of the $j\Delta/B$ in a vertical cross section as denoted by the red line in (a). The arrows in the different panels mark the same location of the current layer.}
  \label{CS}
\end{figure*}

\begin{figure*}
  \centering
  \includegraphics[width=\textwidth]{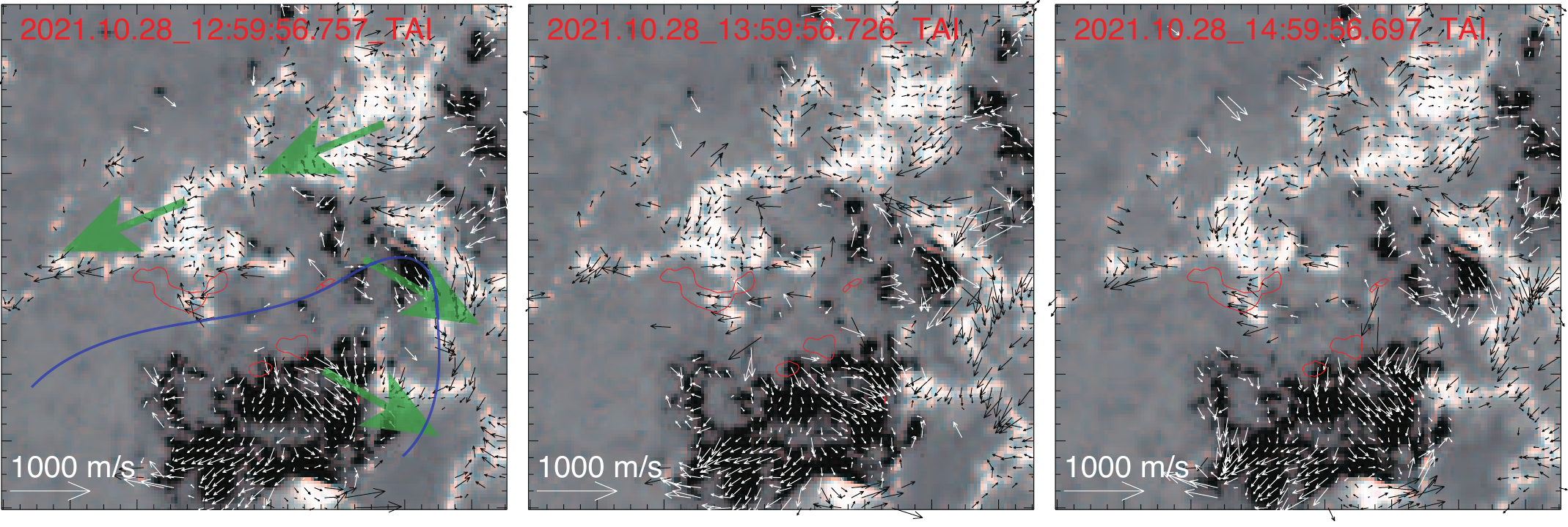}
  \caption{Photospheric velocity field before the eruption onset. The three panels show three different times with cadence of 1 hour. The background shows the magnetogram of $B_z$ and small arrows show the horizontal velocity field (white in the negative polarity and black in the positive polarity).  The red contour lines overlaid show the location of the initial flare ribbons observed by AIA 1600~{\AA} at time of 15:25:51~UT. In the first panel, the blue thick curve depicts the location of the filament (or MFR), and the large green arrows denote the shearing motions of the polarities.}
  \label{flow}
\end{figure*}

To determine more precisely the height of the filament around the eruption onset time, we use a triangulation approach~\citep{Thompson2009} by combining the AIA and EUVI images, and the results of the highest point (which is also roughly the middle point) of the filament are shown in \Fig~\ref{f5}(c). As the selection of the same highest point in both the AIA and EUVI images is done manually, we have performed multiple times of computation and the uncertainty of the estimated heights are shown by the error bars in the plot, which is around $\pm 2$~Mm. For comparison, the heights of the MFR axis at approximately the highest point of the filament for times of 15:00~UT and 15:12~UT derived from the extrapolated NLFFFs are denoted by the red crosses in the plot, which are found to be consistent with the ones derived from the triangulation approach (in particular at time near 15:12~UT). In the slow rise phase until 15:20~UT, the average rising speed of the filament is around $6$~km~s$^{-1}$, while at 15:22:30~UT the filament has already been accelerated to a speed of over $37$~km~s$^{-1}$. As shown in \Fig~\ref{f5}(d), we compute the decay index of the strapping field (i.e., the horizontal component of the potential field, using time of 15:12~UT). The decay index at the height of the filament remains smaller than 1 at least until 15:22~UT when the acceleration of the filament starts. 
 This agrees with our analysis based on the NLFFF reconstruction that the trigger of the eruption is not TI. It also shows that once being triggered, the MFR quickly rises into the TI domain, and after then the eruption should be driven by TI in addition to the reconnection. The peak acceleration, as shown in \Fig~\ref{f3}(c), reaches $2.7$~km~s$^{-2}$ (the real value should be higher, bearing in mind the projected effect). 
A theoretical analysis by~\citet{Vrsnak2008} suggests that without reconnection, an MFR cannot be accelerated in such an impulsive way.

\subsection{Suggestion of reconnection as the initiation mechanism}
As the above analysis shows that the eruption cannot be triggered by TI, here we suggest the reconnection-based mechanism as its trigger with further evidences. Recently using a sufficiently high-accuracy MHD simulation, \citet{JiangC2021} established a fundamental mechanism of solar eruption initiation: a bipolar field driven by slow shearing motion at the photosphere can form an internal current sheet in a quasi-static way, which is followed by fast magnetic reconnection (in the current sheet) that triggers and drives the eruption (see~\citet{JiangC2021} for more details of the process). There are two pieces of evidences suggesting that the studied event follows this reconnection-based scenario. Firstly, the extrapolated magnetic fields before the eruption suggests formation of a current sheet. As shown in \Fig~\ref{CS}, a vertical, thin current layer with a high value of $j\Delta/B \sim 0.2$ (where $\vec j = \crlB$ is the current density and $\Delta$ the grid resolution) is seen at the middle section of the MFR. Note that $j\Delta/B$ is a useful parameter to show current sheet and its value can indicate the thickness of the current layer~\citep[higher value indicates thinner layer, see][]{Jiang2016ApJ, JiangC2021}. From \Fig~\ref{CS}c, it can be seen that the current layer has a thickness of around $2\Delta$. This current layer extends upward from the central part of the flare PIL, and is becoming progressively thinner with time (see \Fig~\ref{CS}d-f). Although this current layer is somewhat thicker than can be regarded as a true current sheet, it is likely to be thinned with further shearing of the field, eventually being a true current sheet until the flare onset.

Secondly, there is continuously photospheric shearing motion before the eruption, which can help to build up the current sheet. \Fig~\ref{flow} shows the photospheric flows as derived by the DAVE4VM method~\citep{Schuck2008} from the time sequence vector magnetograms. As can be seen, although the structure of the flow is complex, the shearing motion, e.g., eastward flow in the north and westward flow in the south, along the PIL under the main branch of the filament (i.e., the core region where the eruption starts) is clearly identified (see the large arrows).

\section{Conclusions}
\label{sec:con}

In this paper, we focus on the initiation mechanism of the first on-disk X-class flare as recorded in solar cycle 25. This flare is accompanied by a filament eruption and results in a moderate CME. The coronal NLFFF reconstructions from the photospheric vector magnetograms reveal that there is a well-defined coronal MFR, with configuration highly consistent with the J-shaped filament, existing for a long period before the flare. During the eruption, the entire filament erupts, which indicates the eruption of the underlying MFR. However, by a quantitative analysis of the extrapolated magnetic field and the observed slow rise phase of the filament immediately before its eruption, we found that the MFR is confined in a height too low to trigger the TI which is often assumed to play the key role for CME initiation of a pre-flare MFR.

On the other hand, the evolution of the current density in the extrapolated magnetic fields shows progressive thinning of a vertical current layer on top of the flare PIL, which hints that a vertical current sheet may form before the eruption. Meanwhile, there is continuously shearing motion along the PIL under the main branch of the filament which can drive the coronal field to form such a current sheet. As such, we suggest that the event follows the reconnection-based initiation mechanism as recently established using a high-accuracy MHD simulation \citep{JiangC2021}: the photosphere shearing motion can build up a current sheet in a quasi-static way, and fast magnetic reconnection in the current sheet triggers the eruption. As accelerated by the reconnection, the filament rises impulsively at the onset of the flare. Then the filament quickly rises into the TI domain and should be further driven by TI in addition to the reconnection, attaining a fast acceleration.

Nevertheless, it should be noted that we still lack a solid indication of the pre-flare current sheet since our analysis is based on the NLFFF extrapolations which has inherent limitations. Indeed, we have never seen in the literature reporting NLFFF extrapolation of coronal current sheet before. The limitations may come from either the force-free assumption or the particular numerical techniques~\citep[e.g., see][]{Low2013,  Yeates2022}. For example, the force-free assumption might fail if there is a non-force-free current sheet in which the Lorentz force is balanced by gas pressure gradient, and since the force-free field assumes a smooth solution while a true current sheet is a discontinuity, which constitutes a weak solution of the MHD equation. It is also possible that a current sheet exists, but the grid resolution of the extrapolation is not sufficiently high to capture the current sheet.
Future investigation of this event will be carried out by high-resolution data-driven MHD simulations~\citep{JiangC2022} to study how a current sheet is built up as driven by the photospheric motions along with the presence of the MFR.

\begin{acknowledgements} This work is supported by National Natural Science Foundation of China (NSFC) U2031108, Guangdong Basic and Applied Basic Research Foundation (2021A1515011430), and Yunnan Key Laboratory of Solar Physics and Space Science under the number YNSPCC202213. C.J. acknowledges
support by NSFC 42174200, Guangdong Basic and Applied Basic Research Foundation (2023B1515040021), and Shenzhen Science and Technology Program (Grant No. RCJC20210609104422048). The computational work of the NLFFF
extrapolations was carried out on TianHe-1(A), National Supercomputer
Center in Tianjin, China. Data from observations are courtesy of
NASA/SDO and Cerro Tololo Inter-American Observatory.
We are very grateful to the reviewer for helpful comments and suggestions that improved the paper.
\end{acknowledgements}

\bibliographystyle{aa}

\begin{thebibliography}{58}
\expandafter\ifx\csname natexlab\endcsname\relax\def\natexlab#1{#1}\fi

\bibitem[{Alt {et~al.}(2021)Alt, Myers, Ji, Jara-Almonte, Yoo, Bose, Goodman,
  Yamada, Kliem, \& Savcheva}]{Alt2021}
Alt, A., Myers, C.~E., Ji, H., {et~al.} 2021, The Astrophysical Journal
  (Online), 908

\bibitem[{Amari {et~al.}(2014)Amari, Canou, \& Aly}]{Amari2014nat}
Amari, T., Canou, A., \& Aly, J.~J. 2014, Nature, 514, 465

\bibitem[{{Aulanier} {et~al.}(2010){Aulanier}, {T{\"o}r{\"o}k}, {D{\'e}moulin},
  \& {DeLuca}}]{Aulanier2010}
{Aulanier}, G., {T{\"o}r{\"o}k}, T., {D{\'e}moulin}, P., \& {DeLuca}, E.~E.
  2010, \apj, 708, 314

\bibitem[{{Berger} \& {Prior}(2006)}]{Berger2006}
{Berger}, M.~A. \& {Prior}, C. 2006, Journal of Physics A Mathematical General,
  39, 8321

\bibitem[{{Bobra} {et~al.}(2014){Bobra}, {Sun}, {Hoeksema}, {Turmon}, {Liu},
  {Hayashi}, {Barnes}, \& {Leka}}]{Bobra2014}
{Bobra}, M.~G., {Sun}, X., {Hoeksema}, J.~T., {et~al.} 2014, \solphys, 289,
  3549

\bibitem[{{Chen}(1989)}]{ChenJ1989}
{Chen}, J. 1989, \apj, 338, 453

\bibitem[{Chen(2011)}]{ChenP2011}
Chen, P.~F. 2011, Living Reviews in Solar Physics, 8, 1

\bibitem[{Cheng {et~al.}(2017)Cheng, Guo, \& Ding}]{ChengX2017}
Cheng, X., Guo, Y., \& Ding, M. 2017, Science China Earth Sciences, 60, 1383

\bibitem[{{D{\'e}moulin} \& {Aulanier}(2010)}]{Demoulin2010}
{D{\'e}moulin}, P. \& {Aulanier}, G. 2010, \apj, 718, 1388

\bibitem[{{DeRosa} {et~al.}(2009){DeRosa}, {Schrijver}, {Barnes}, {Leka},
  {Lites}, {Aschwanden}, {Amari}, {Canou}, {McTiernan}, {R{\'e}gnier},
  {Thalmann}, {Valori}, {Wheatland}, {Wiegelmann}, {Cheung}, {Conlon},
  {Fuhrmann}, {Inhester}, \& {Tadesse}}]{DeRosa2009}
{DeRosa}, M.~L., {Schrijver}, C.~J., {Barnes}, G., {et~al.} 2009, \apj, 696,
  1780

\bibitem[{{DeRosa} {et~al.}(2015){DeRosa}, {Wheatland}, {Leka}, {Barnes},
  {Amari}, {Canou}, {Gilchrist}, {Thalmann}, {Valori}, {Wiegelmann},
  {Schrijver}, {Malanushenko}, {Sun}, \& {R{\'e}gnier}}]{DeRosa2015}
{DeRosa}, M.~L., {Wheatland}, M.~S., {Leka}, K.~D., {et~al.} 2015, \apj, 811,
  107

\bibitem[{Duan {et~al.}(2019)Duan, Jiang, He, Feng, Zou, \& Cui}]{DuanA2019}
Duan, A., Jiang, C., He, W., {et~al.} 2019, The Astrophysical Journal, 884, 73

\bibitem[{{Duan} {et~al.}(2017){Duan}, {Jiang}, {Hu}, {Zhang}, {Gary}, {Wu}, \&
  {Cao}}]{DuanA2017}
{Duan}, A., {Jiang}, C., {Hu}, Q., {et~al.} 2017, \apj, 842, 119

\bibitem[{{Duan} {et~al.}(2021{\natexlab{a}}){Duan}, {Jiang}, {Zhou}, {Feng},
  \& {Cui}}]{DuanA2021apjl}
{Duan}, A., {Jiang}, C., {Zhou}, Z., {Feng}, X., \& {Cui}, J.
  2021{\natexlab{a}}, \apjl, 907, L23

\bibitem[{{Duan} {et~al.}(2021{\natexlab{b}}){Duan}, {Jiang}, {Zou}, {Feng}, \&
  {Cui}}]{DuanA2021apj}
{Duan}, A., {Jiang}, C., {Zou}, P., {Feng}, X., \& {Cui}, J.
  2021{\natexlab{b}}, \apj, 906, 45

\bibitem[{Duan {et~al.}(2022)Duan, Jiang, Guo, Feng, \& Cui}]{DuanA2022AA}
Duan, A.~Y., Jiang, C.~W., Guo, Y., Feng, X.~S., \& Cui, J. 2022, Astronomy \&
  Astrophysics

\bibitem[{Fan(2010)}]{Fan2010}
Fan, Y. 2010, \apj, 719, 728

\bibitem[{{Fan} \& {Gibson}(2007)}]{Fan2007}
{Fan}, Y. \& {Gibson}, S.~E. 2007, \apj, 668, 1232

\bibitem[{{Forbes} {et~al.}(2006){Forbes}, {Linker}, {Chen}, {Cid}, {K{\'o}ta},
  {Lee}, {Mann}, {Miki{\'c}}, {Potgieter}, {Schmidt}, {Siscoe}, {Vainio},
  {Antiochos}, \& {Riley}}]{Forbes2006}
{Forbes}, T.~G., {Linker}, J.~A., {Chen}, J., {et~al.} 2006, \ssr, 123, 251

\bibitem[{Guo {et~al.}(2017)Guo, Cheng, \& Ding}]{GuoY2017}
Guo, Y., Cheng, X., \& Ding, M. 2017, Science China Earth Sciences, 60, 1408

\bibitem[{{Guo} {et~al.}(2010){Guo}, {Schmieder}, {D{\'e}moulin}, {Wiegelmann},
  {Aulanier}, {T{\"o}r{\"o}k}, \& {Bommier}}]{Guo2010}
{Guo}, Y., {Schmieder}, B., {D{\'e}moulin}, P., {et~al.} 2010, \apj, 714, 343

\bibitem[{{Hood} \& {Priest}(1981)}]{Hood1981}
{Hood}, A.~W. \& {Priest}, E.~R. 1981, Geophysical and Astrophysical Fluid
  Dynamics, 17, 297

\bibitem[{Inoue {et~al.}(2014)Inoue, Hayashi, Magara, Choe, \&
  Park}]{Inoue2014}
Inoue, S., Hayashi, K., Magara, T., Choe, G.~S., \& Park, Y.~D. 2014, The
  Astrophysical Journal, 788, 182

\bibitem[{Inoue {et~al.}(2018)Inoue, Kusano, B{\"u}chner, \&
  Sk{\'a}la}]{Inoue2018}
Inoue, S., Kusano, K., B{\"u}chner, J., \& Sk{\'a}la, J. 2018, Nature
  communications, 9, 174

\bibitem[{{Jiang} \& {Feng}(2013)}]{Jiang2013NLFFF}
{Jiang}, C. \& {Feng}, X. 2013, \apj, 769, 144

\bibitem[{{Jiang} \& {Feng}(2014)}]{Jiang2014Prep}
{Jiang}, C. \& {Feng}, X. 2014, \solphys, 289, 63

\bibitem[{Jiang {et~al.}(2022)Jiang, Feng, Guo, \& Hu}]{JiangC2022}
Jiang, C., Feng, X., Guo, Y., \& Hu, Q. 2022, The Innovation, 3, 100236

\bibitem[{Jiang {et~al.}(2021)Jiang, Feng, Liu, Yan, Hu, Moore, Duan, Cui, Zuo,
  Wang, \& Wei}]{JiangC2021}
Jiang, C., Feng, X., Liu, R., {et~al.} 2021, Nature Astronomy, 5, 1126

\bibitem[{Jiang {et~al.}(2016)Jiang, Wu, Yurchyshyn, Wang, Feng, \&
  Hu}]{Jiang2016ApJ}
Jiang, C., Wu, S.~T., Yurchyshyn, V.~B., {et~al.} 2016, \apj, 828, 62

\bibitem[{{Jiang} \& {Feng}(2012)}]{Jiang2012apj}
{Jiang}, C.~W. \& {Feng}, X.~S. 2012, \apj, 749, 135

\bibitem[{Jiang {et~al.}(2010)Jiang, Feng, Zhang, \& Zhong}]{Jiang2010}
Jiang, C.~W., Feng, X.~S., Zhang, J., \& Zhong, D.~K. 2010, \solphys, 267, 463

\bibitem[{{Kliem} \& {T{\"o}r{\"o}k}(2006)}]{Kliem2006}
{Kliem}, B. \& {T{\"o}r{\"o}k}, T. 2006, Physical Review Letters, 96, 255002

\bibitem[{{Kuperus} \& {Raadu}(1974)}]{Kuperus1974}
{Kuperus}, M. \& {Raadu}, M.~A. 1974, \aap, 31, 189

\bibitem[{Kusano {et~al.}(2020)Kusano, Iju, Bamba, \& Inoue}]{Kusano2020}
Kusano, K., Iju, T., Bamba, Y., \& Inoue, S. 2020, Science, 369, 587

\bibitem[{Lemen {et~al.}(2012)Lemen, Title, Akin, Boerner, Chou, Drake, Duncan,
  Edwards, Friedlaender, Heyman, Hurlburt, Katz, Kushner, Levay, Lindgren,
  Mathur, McFeaters, Mitchell, Rehse, Schrijver, Springer, Stern, Tarbell,
  Wuelser, Wolfson, Yanari, Bookbinder, Cheimets, Caldwell, Deluca, Gates,
  Golub, Park, Podgorski, Bush, Scherrer, Gummin, Smith, Auker, Jerram, Pool,
  Soufli, Windt, Beardsley, Clapp, Lang, \& Waltham}]{Lemen2012}
Lemen, J.~R., Title, A.~M., Akin, D.~J., {et~al.} 2012, Solar Physics, 275, 17

\bibitem[{Liu {et~al.}(2016)Liu, Kliem, Titov, Chen, Wang, Wang, Liu, Xu, \&
  Wiegelmann}]{LiuR2016}
Liu, R., Kliem, B., Titov, V.~S., {et~al.} 2016, The Astrophysical Journal,
  818, 148

\bibitem[{{Low}(2013)}]{Low2013}
{Low}, B.~C. 2013, \apj, 768, 7

\bibitem[{{Malanushenko} {et~al.}(2014){Malanushenko}, {Schrijver}, {DeRosa},
  \& {Wheatland}}]{Malanushenko2014}
{Malanushenko}, A., {Schrijver}, C.~J., {DeRosa}, M.~L., \& {Wheatland}, M.~S.
  2014, \apj, 783, 102

\bibitem[{McCauley {et~al.}(2015)McCauley, Su, Schanche, Evans, Su, McKillop,
  \& Reeves}]{McCauley2015}
McCauley, P.~I., Su, Y.~N., Schanche, N., {et~al.} 2015, Solar Physics, 290,
  1703

\bibitem[{{Myers} {et~al.}(2015){Myers}, {Yamada}, {Ji}, {Yoo}, {Fox},
  {Jara-Almonte}, {Savcheva}, \& {Deluca}}]{Myers2015}
{Myers}, C.~E., {Yamada}, M., {Ji}, H., {et~al.} 2015, \nat, 528, 526

\bibitem[{Pesnell {et~al.}(2012)Pesnell, Thompson, \& Chamberlin}]{Pesnell2012}
Pesnell, W.~D., Thompson, B.~J., \& Chamberlin, P.~C. 2012, Solar Physics, 275,
  3

\bibitem[{{Schou} {et~al.}(2012){Schou}, {Scherrer}, {Bush}, {Wachter},
  {Couvidat}, {Rabello-Soares}, {Bogart}, {Hoeksema}, {Liu}, {Duvall}, {Akin},
  {Allard}, {Miles}, {Rairden}, {Shine}, {Tarbell}, {Title}, {Wolfson},
  {Elmore}, {Norton}, \& {Tomczyk}}]{Schou2012}
{Schou}, J., {Scherrer}, P.~H., {Bush}, R.~I., {et~al.} 2012, \solphys, 275,
  229

\bibitem[{{Schrijver} {et~al.}(2006){Schrijver}, {De Rosa}, {Metcalf}, {Liu},
  {McTiernan}, {R{\'e}gnier}, {Valori}, {Wheatland}, \&
  {Wiegelmann}}]{Schrijver2006}
{Schrijver}, C.~J., {De Rosa}, M.~L., {Metcalf}, T.~R., {et~al.} 2006,
  \solphys, 235, 161

\bibitem[{{Schrijver} {et~al.}(2008){Schrijver}, {DeRosa}, {Metcalf}, {Barnes},
  {Lites}, {Tarbell}, {McTiernan}, {Valori}, {Wiegelmann}, {Wheatland},
  {Amari}, {Aulanier}, {D{\'e}moulin}, {Fuhrmann}, {Kusano}, {R{\'e}gnier}, \&
  {Thalmann}}]{Schrijver2008a}
{Schrijver}, C.~J., {DeRosa}, M.~L., {Metcalf}, T., {et~al.} 2008, \apj, 675,
  1637

\bibitem[{Schuck(2008)}]{Schuck2008}
Schuck, P.~W. 2008, The Astrophysical Journal, 683, 1134

\bibitem[{{Shibata} \& {Magara}(2011)}]{Shibata2011}
{Shibata}, K. \& {Magara}, T. 2011, Living Reviews in Solar Physics, 8, 6

\bibitem[{Thompson(2009)}]{Thompson2009}
Thompson, W. 2009, Icarus, 200, 351

\bibitem[{{Titov} \& {D{\'e}moulin}(1999)}]{Titov1999}
{Titov}, V.~S. \& {D{\'e}moulin}, P. 1999, \aap, 351, 707

\bibitem[{{T{\"o}r{\"o}k} {et~al.}(2010){T{\"o}r{\"o}k}, {Berger}, \&
  {Kliem}}]{Torok2010}
{T{\"o}r{\"o}k}, T., {Berger}, M.~A., \& {Kliem}, B. 2010, \aap, 516, A49

\bibitem[{{T{\"o}r{\"o}k} \& {Kliem}(2005)}]{Torok2005}
{T{\"o}r{\"o}k}, T. \& {Kliem}, B. 2005, \apjl, 630, L97

\bibitem[{{T{\"o}r{\"o}k} \& {Kliem}(2007)}]{Torok2007}
{T{\"o}r{\"o}k}, T. \& {Kliem}, B. 2007, Astronomische Nachrichten, 328, 743

\bibitem[{{T{\"o}r{\"o}k} {et~al.}(2004){T{\"o}r{\"o}k}, {Kliem}, \&
  {Titov}}]{Torok2004}
{T{\"o}r{\"o}k}, T., {Kliem}, B., \& {Titov}, V.~S. 2004, \aap, 413, L27

\bibitem[{{Vr{\v{s}}nak}(2008)}]{Vrsnak2008}
{Vr{\v{s}}nak}, B. 2008, Annales Geophysicae, 26, 3089

\bibitem[{{Yeates}(2022)}]{Yeates2022}
{Yeates}, A.~R. 2022, Geophysical and Astrophysical Fluid Dynamics, 116, 305

\bibitem[{Zou {et~al.}(2019)Zou, Jiang, Feng, Zuo, Wang, \& Wei}]{ZouP2019}
Zou, P., Jiang, C., Feng, X., {et~al.} 2019, The Astrophysical Journal, 870, 97

\bibitem[{Zou {et~al.}(2020)Zou, Jiang, Wei, Feng, Zuo, \& Wang}]{ZouP2020}
Zou, P., Jiang, C., Wei, F., {et~al.} 2020, The Astrophysical Journal, 890, 10

\bibitem[{Zuccarello {et~al.}(2015)Zuccarello, Aulanier, \&
  Gilchrist}]{Zuccarello2015}
Zuccarello, F.~P., Aulanier, G., \& Gilchrist, S.~A. 2015, The Astrophysical
  Journal, 814, 126

\bibitem[{Zuccarello {et~al.}(2016)Zuccarello, Aulanier, \&
  Gilchrist}]{Zuccarello2016}
Zuccarello, F.~P., Aulanier, G., \& Gilchrist, S.~A. 2016, The Astrophysical
  Journal Letters, 821, L23

\end{thebibliography}

\end{document}